%% file: X_paper_fermilab.tex
\documentclass[prd,twocolumn,showpacs,amsmath,amssymb]{revtex4}

\usepackage{graphicx}
\usepackage{dcolumn}
\usepackage{bm}

\begin{document}

\hspace{5.2in} \mbox{FERMILAB-Pub-04/061-E}

\title{\boldmath \bf Observation and Properties of the $X(3872)$ 
Decaying to $J/\psi \, \pi^{+} \pi^{-}$ \\
 in $p\bar{p}$ Collisions at $\sqrt{s}$= 1.96 TeV}



\input list_of_authors_r2.tex

           
\begin{abstract}
We report the observation of the $X(3872)$ in the $J/\psi
\, \pi^{+} \pi^{-}$ channel, with $J/\psi$ decaying to $\mu^{+} \mu^{-}$, in
$p\bar{p}$ collisions at $\sqrt{s}$= 1.96 TeV.  Using approximately
230~pb$^{-1}$ of data collected with the Run II D\O\ detector, we
observe 522 $\pm$ 100 $X(3872)$ candidates.  The mass difference 
between the $X(3872)$ state and the $J/\psi$ is measured
to be $774.9~
\pm$ $ 3.1$~(stat)~$\pm$ $3.0$~(syst) MeV/$c^{2}$.  We have investigated
the production and decay characteristics of the $X(3872)$,
and find them to be similar to those of the $\psi(2S)$ state.

\end{abstract}
 
\pacs{14.40.Gx, 14.40.Lb, 13.25.Gv, 13.25.Jx, 13.85.Ni}

\maketitle

\newpage

 A new particle, the $X(3872)$, was recently discovered by the
 Belle Collaboration~\cite{Belle} in the decay mode $B^{\pm} \rightarrow
 X K^{\pm}$ ($X \rightarrow J/\psi \, \pi^{+} \pi^{-}$) where the mass of
 the $X(3872)$ was measured to be 3872.0 $\pm$ 0.6~(stat)
 $\pm$ 0.5 (syst) MeV/$c^{2}$.  The existence of the $X(3872)$ state
 has been confirmed (also decaying in the $J/\psi \, \pi^{+} \pi^{-}$ mode)
 in $p \bar{p}$ collisions by the CDF Collaboration~\cite{CDF}. At
 this time, it is still unclear whether this particle is a $c\bar{c}$ state,
 or a more complex object.  See for example ~\cite{theory2,theory,Quigg:2004nv,Tornqvist}.

The charmonium state $\psi(2S)$, with mass $m_{\psi(2S)}$= $3685.96 \pm
0.09$~MeV/$c^2$~\cite{PDG}, has the same decay mode, providing a good 
benchmark for comparison with
the $X(3872)$.  The $\psi(2S)$ mesons produced in $p {\bar p}$ collisions can originate either from decays of $B$ hadrons or from direct production.
The $\psi(2S)$ mesons from $B$
decays have longer
effective decay lengths and tend to be less isolated than directly
produced $\psi(2S)$ mesons ~\cite{jpsi_prod}.

We examine the production rate of the $X(3872)$ relative to $\psi(2S)$
as a function of the transverse momentum with respect to the beam axis ($p_T$),
isolation and decay length, as well as a function of rapidity 
($y=\frac{1}{2} \, {\rm log} \, \frac{E+P_{L}}{E-P_{L}}$, 
where $E$ is the energy and $P_{L}$ is the longitudinal momentum with respect to the beam axis),
to determine whether the production characteristics of the
$X(3872)$ are similar to those of the $\psi(2S)$.  
We also compare the angular decay distributions of the $\pi^+
\pi^-$ and $\mu^+ \mu^-$ systems in $X(3872)$ decays with those
from $\psi(2S)$, to check for any differences in helicities of these two 
states.



The data set used in this Letter was collected in $p\bar{p}$ collisions
at $\sqrt{s}$=1.96 TeV between April 2002 and January 2004, and
corresponds to an integrated luminosity of approximately 230~pb$^{-1}$.
The D\O\ detector is
described elsewhere~\cite{d0det}.  The components
most important to this analysis include the vertex, central tracking and
muon systems.  The D\O\ tracking system consists of a
silicon microstrip tracker (SMT) and a central fiber tracker (CFT),
both within a 2 T solenoidal magnetic field.  

The SMT has approximately $800,000$ individual strips, with typical
pitch of 50--80~$\mu$m, and a design optimized for tracking and
vertexing over the range $|\eta| < 3$, where $\eta =
-\ln[\tan(\theta/2)]$ is the pseudorapidity and $\theta$ is the polar
angle measured relative to the proton beam direction.  The system has a
six-barrel longitudinal structure, each with a set of four layers
arranged axially around the beam pipe, and interspersed with 16 radial
disks.  The system provides a resolution, in the plane transverse to the
beam axis, for the distance of closest
approach of a charged particle relative to the primary vertex of
$\approx 50$~$\mu$m for tracks with $p_T$ $\approx$ 1 GeV/$c$, 
improving asymptotically to
15~$\mu$m for tracks with $p_T$ $\ge 10$~GeV/$c$.

The CFT comprises eight thin coaxial barrels, each
supporting two doublets of overlapping scintillating fibers
of 0.835~mm diameter, one doublet being parallel to the 
collision axis, and the other alternating by $\pm 3^{\circ}$
to provide information along the beam axis.
  
The muon system is located outside the calorimeters,
and consists of a layer of tracking detectors and 
scintillation trigger counters in front of 1.8~T toroidal magnets,
followed by two similar layers behind the toroids.
Tracking in the muon system in the range 
$|\eta| < 1$ relies on 10~cm wide drift
tubes~\cite{pdts}, while 1~cm mini-drift tubes are
used for $1 < |\eta| < 2$.

$J/\psi \rightarrow \mu^{+} \mu^{-}$ decays are selected by
triggering on dimuons using a three-tier trigger system. The first
trigger level uses hardware to form roads defined by hits in
two layers of the muon scintillator system.  The second trigger level
uses digital signal processors to form track stubs defined by hits in
the muon drift-chamber and muon scintillator systems.  
The third level comprises a farm
of computer processors with access to the entire event. Events passing
the third-level trigger are recorded for analysis. 


\begin{figure}[htb]
\begin{center}
\includegraphics[width=3.5in]{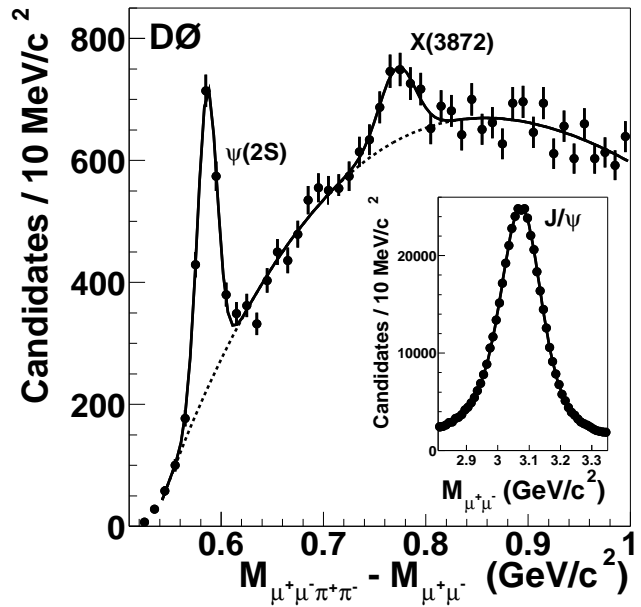}
\end{center}
\caption{$\Delta M = M(\mu^{+} \mu^{-} \pi^{+}
\pi^{-}) - M(\mu^{+}\mu^{-})$ for all candidates satisfying 
the selection requirements.  The solid curve is a fit to the data and
the dashed curve represents the background under each peak.  The insert 
shows the mass distribution of the $J/\psi$ candidates used in the analysis.}
\label{fig:with_X}
\end{figure}




Muons are identified by extrapolating charged particle tracks from
the central tracking system that match with muon track segments
formed from hits in the muon system.  Oppositely charged muons are
combined to form $J/\psi$ candidates, which are then combined with
two oppositely charged particles assumed to be pions.
At least two of these four tracks are required to have at
least one hit in the SMT. To reduce background from combinatorics, events
are required to satisfy the following selection criteria.
An event is required to have less than 100 tracks.
$J/\psi$
candidates are selected by requiring the invariant mass of the
$\mu^+\mu^-$ system to be between 2.80 and 3.35 GeV/$c^2$, and the transverse
momentum with respect to the beam axis ($p_{T}$)
of the $J/\psi$ is required to be greater than 5~GeV/$c$.  In addition,
the $p_{T}$ of each of the two pions must be greater than 0.7 GeV/$c$, and the
spatial separation, $\Delta R$, between the momentum vector of the
$J/\psi \, \pi^{+} \pi^{-}$ system and each pion momentum vector is
required to have $\Delta R$ $<$ 0.4, where  $\Delta R$ is defined as
$\sqrt{(\Delta\phi)^2 + (\Delta\eta)^2}$, with $\phi$ being the azimuthal
angle. The invariant mass of the two pions, $M(\pi^+\pi^-)$, is
required to be greater than 0.52 GeV/$c^2$, and the $\chi^2$ of a fit to
the $\mu^{+} \mu^{-}
\pi^+ \pi^-$ vertex is required to be less than 16 (for five
degrees of freedom).


Figure \ref{fig:with_X} shows the distribution in the mass difference 
$\Delta M = M(\mu^{+} \mu^{-} \pi^{+}
\pi^{-}) - M(\mu^{+}\mu^{-})$, after all selections. 
Superimposed is a fit to the data, where Gaussians are
used to describe the 
$\psi(2S)$ and the $X(3872)$, and a third-order polynomial is used
to account for the background.  
The fitted width of the $X(3872)$ peak is $17 \pm 3$~MeV/$c^2$, 
which is consistent with detector resolution.  
The results of the Gaussian fit yield $522 \pm 100$ $X(3872)$ candidates
with $\Delta M = 771.9 \pm 3.1$~(stat)~MeV/$c^2$.
The yield of the $X(3872)$ is dependent on the value of
the $M(\pi^+\pi^-)$ requirement.  
When we tighten the requirement on $M(\pi^+\pi^-)$ to be greater 
than 0.60 GeV/$c^2$, 0.65 GeV/$c^2$ and 0.70 GeV/$c^2$ and fix the mean of the $X(3872)$ 
to its nominal value, we obtain yields of 433$\pm$98, 336$\pm$78 and 170$\pm$54 events, respectively.

The position of the $J/\psi$ mass in Fig. \ref{fig:with_X} is shifted
relative to the currently accepted value~\cite{PDG}. 
This is within the uncertainties of our momentum scale, 
but we correct the measured $\Delta M$ by
the ratio of the currently accepted mass difference between the
$\psi(2S)$ and the $J/\psi$~\cite{PDG} and our corresponding measured mass
difference.  We assign a 100\% systematic uncertainty to this mass
scale correction, giving a final measurement of 
$\Delta M = 774.9 \pm
3.1$~(stat) $\pm$ 3.0 (syst)~MeV/$c^2$. We investigated
systematic effects by using different background parameterizations and bin
widths in our fits, and found the 
largest change in the central value of the mass difference
to be 0.6~MeV/$c^2$.

\begin{figure}[htb]
\begin{center}
\includegraphics[height=3.2in]{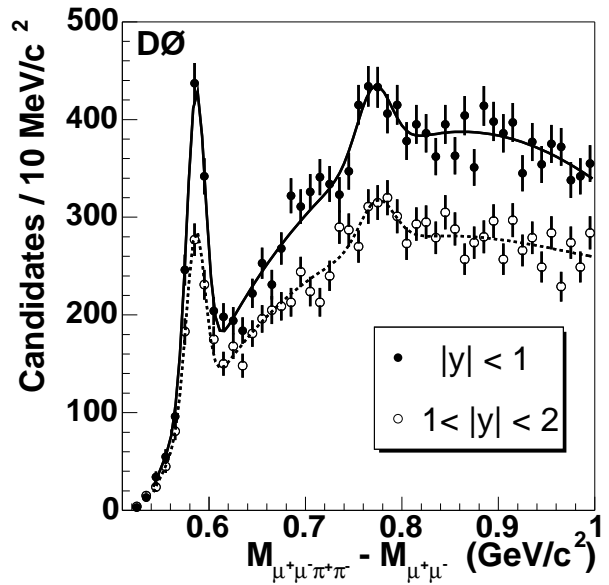}
\end{center}
\caption{$\Delta M$ distribution for two ranges of rapidity
of the $\mu^+ \mu^- \pi^+ \pi^-$ system.}
\label{fig:y}
\end{figure}

To investigate the characteristics of the $X(3872)$ state, we
study its production and decay properties, and
compare the signal yields of the $X(3872)$ to the $\psi(2S)$.  For
example, Fig.~\ref{fig:y} shows the $\Delta M$ distribution in two
separate regions of rapidity of the $X(3872)$.  
The data is also separated into regions of
effective proper decay length, transverse momentum of
the $X(3872)$ candidate, isolation, and decay angle.

The effective proper decay length, $dl$, is defined as
the distance in the transverse plane 
from the primary vertex to the decay vertex of the
$J/\psi$ scaled by the mass of the $\mu^+ \mu^- \pi^+ \pi^-$ system
divided by the $\mu^+ \mu^- \pi^+ \pi^-$ system $p_T$.

The isolation 
is defined as the ratio of the $X(3872)$ momentum to the sum
of the momentum of the $X(3872)$ and the momenta of all other reconstructed 
charged particles within a
cone of radius $\Delta R = 0.5$, about the direction of the
$X(3872)$ momentum.


The helicity of the $\pi^+\pi^- \,(\mu^+\mu^-)$ system can be inferred by 
boosting one of the pions (or muons) and the $X(3872)$ into the dipion 
(dimuon) rest frame, and measuring the angle $\theta_\pi (\theta_\mu)$ between them. The cosine of this angle is used for the 
comparison between the $X(3872)$ and $\psi$(2S).

\begin{figure}[htb]
\begin{center}
\includegraphics[height=3.2in]{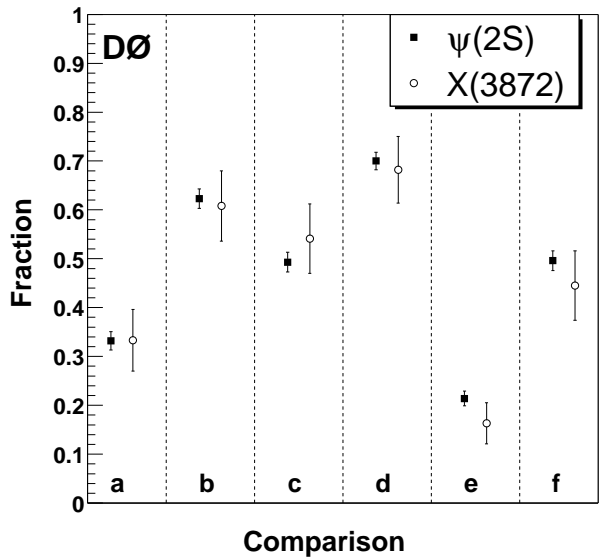}
\end{center}
\caption{Comparison of event-yield fractions for $X(3872)$ and $\psi(2S)$ in the regions: 
(a) $p_T$($J/\psi \, \pi^+ \pi^-) >15$ GeV/$c$; (b) $|y|$ of $J/\psi \, \pi^+ 
\pi^- < 1$;  (c)  cos($\theta_{\pi})$ $< 0.4$; (d) effective proper decay 
length, $dl$ $< 0.01$ 
 cm; (e) isolation = 1; (f) cos($\theta_{\mu})$ $< 0.4$.   }
\label{fig:fractions}
\end{figure}

The widths of the $\psi(2S)$ and $X(3872)$ are fixed to the values
obtained in the fit to the full sample, and the fitted number of
$\psi(2S)$ and $X(3872)$ candidates in each region of chosen variables
is given in Table~\ref{tab:table1}.
The data in all regions are well represented by our choice of
fitting function, with fit probabilities all greater than 15\%.
In Fig.~\ref{fig:fractions}, we present the results of Table~\ref{tab:table1} 
in a graphical form.
For example, we compare the event yield fraction:
$$ f_{X(3872)}= \frac{N(X(3872)) ~ |y| < 1}
                  {N(X(3872)) ~ |y| < 2} , $$ 
to that for $\psi(2S)$ : 
$$ f_{\psi(2S)} = \frac{ N(\psi(2S)) ~ |y| < 1}
                       { N(\psi(2S)) ~ |y| < 2} .$$

\begin{table}
\caption{\label{tab:table1}
Number of $\psi(2S)$ and $X(3872)$ candidates
for different ranges of variables.  The fitted widths of the
$X(3872)$ and $\psi(2S)$ are not constrained in the initial selection, but
in the different ranges they are fixed to the values 
obtained from the full sample.
Apart from the full sample, uncertainties are only from the
normalization of the fitted Gaussian function and do not include small
contributions from uncertainties in the background.}
\begin{ruledtabular}
\begin{tabular}{lcc}
Regions & Number & Number \\
     &        of $\psi(2S)$ &        of $X(3872)$ \\
\hline
Initial selection & 1192  $\pm$ 55 & 522 $\pm$ 100 \\ \hline
(a) ~$p_T(J/\psi \, \pi^+ \pi^-) > 15$~GeV/$c$  & 396 $\pm$ 26 & 179 $\pm$ 39 \\
~~~~~~$p_T(J/\psi \, \pi^+ \pi^-) \le 15$~GeV/$c$ & 796 $\pm$ 41 & 358 $\pm$ 64 \\ \hline
(b) ~$|y| < 1$  & 741 $\pm$ 37 & 316 $\pm$ 57 \\
~~~~~~$1 \le |y| \le 2$ & 449 $\pm$ 31 & 204 $\pm$ 49 \\ \hline
(c) ~cos($\theta_{\pi})$ $<$ 0.4  & 589 $\pm$ 34 & 288 $\pm$ 53 \\
$\,$~~~~~cos($\theta_{\pi})$ $\ge$ 0.4 & 606 $\pm$ 34 & 244 $\pm$ 53\\ \hline
(d) ~$dl$ $<$ 0.01 cm  & 838 $\pm$ 41 & 351 $\pm$ 66\\
~~~~~~$dl$ $\ge$ 0.01 cm & 359 $\pm$ 26 & 164 $\pm$ 41 \\ \hline
(e) ~isolation  = 1  & 257 $\pm$ 20 & 85 $\pm$ 22\\
~~~~~~isolation  $<$ 1 & 942 $\pm$ 44 & 438 $\pm$ 72\\ \hline
(f) ~cos($\theta_{\mu})$ $<$ 0.4  & 593 $\pm$ 33 & 232 $\pm$ 46 \\
$\,$~~~~~cos($\theta_{\mu})$ $\ge$ 0.4 & 602 $\pm$ 35 & 288 $\pm$ 60\\
\end{tabular}
\end{ruledtabular}
\end{table}

From the ratios of isolation and decay lengths, it is apparent that
the production of the $X(3872)$ has a similar mixture of prompt
and long-lived contributions as the $\psi(2S)$.

In summary, we observe the $X(3872)$ particle in approximately
230~pb$^{-1}$ of data with a significance of 5.2 standard deviations.
The mass difference between the $X(3872)$ particle and the
$J/\psi$ is $\Delta M = 774.9 \pm 3.1$~(stat) $\pm$
3.0~(syst)~MeV/$c^2$.  When the data are separated according to
production and decay variables, we find no significant
differences between the $X(3872)$ and
the $c \bar{c}$ state $\psi(2S)$.

We thank Estia Eichten, Ken Lane, and Chris Quigg, for fruitful
discussions. 
\input acknowledgement_paragraph_r2.tex


\end{document}

%% file: list_of_authors_r2.tex
%
\author{                                                                      
V.M.~Abazov,$^{32}$                                                           
B.~Abbott,$^{69}$                                                             
M.~Abolins,$^{60}$                                                            
B.S.~Acharya,$^{26}$                                                          
D.L.~Adams,$^{67}$                                                            
M.~Adams,$^{47}$                                                              
T.~Adams,$^{45}$                                                              
M.~Agelou,$^{16}$                                                             
J.-L.~Agram,$^{17}$                                                           
S.N.~Ahmed,$^{31}$
% ...
\\                                                                            
\vskip 0.30cm                                                                 
\centerline{(D\O\ Collaboration)}                                             
\vskip 0.30cm                                                                 
}                                                                             
\address{                                                                     
\centerline{$^{1}$Universidad de Buenos Aires, Buenos Aires, Argentina}       
\centerline{$^{2}$LAFEX, Centro Brasileiro de Pesquisas F{\'\i}sicas,         
                  Rio de Janeiro, Brazil}                                     
\centerline{$^{3}$Universidade do Estado do Rio de Janeiro,                   
                  Rio de Janeiro, Brazil}                                     
\centerline{$^{4}$Instituto de F\'{\i}sica Te\'orica, Universidade            
                  Estadual Paulista, S\~ao Paulo, Brazil}                     
\centerline{$^{5}$University of Alberta and Simon Fraser University,          
                  Canada}                                                     
\centerline{$^{6}$Institute of High Energy Physics, Beijing,                  
                  People's Republic of China}                                 
\centerline{$^{7}$Universidad de los Andes, Bogot\'{a}, Colombia}             
\centerline{$^{8}$Charles University, Center for Particle Physics,            
                  Prague, Czech Republic}                                     
\centerline{$^{9}$Czech Technical University, Prague, Czech Republic}         
\centerline{$^{10}$Institute of Physics, Academy of Sciences, Center          
                  for Particle Physics, Prague, Czech Republic}               
\centerline{$^{11}$Universidad San Francisco de Quito, Quito, Ecuador}        
\centerline{$^{12}$Laboratoire de Physique Subatomique et de Cosmologie,      
                  IN2P3-CNRS, Universite de Grenoble 1, Grenoble, France}     
\centerline{$^{13}$CPPM, IN2P3-CNRS, Universit\'e de la M\'editerran\'ee,     
                  Marseille, France}                                          
\centerline{$^{14}$Laboratoire de l'Acc\'el\'erateur Lin\'eaire,              
                  IN2P3-CNRS, Orsay, France}                                  
\centerline{$^{15}$LPNHE, Universit\'es Paris VI and VII, IN2P3-CNRS,         
                  Paris, France}                                              
\centerline{$^{16}$DAPNIA/Service de Physique des Particules, CEA, Saclay,    
                  France}                                                     
\centerline{$^{17}$IReS, IN2P3-CNRS, Univ. Louis Pasteur Strasbourg,          
                   and Univ. de Haute Alsace, France}                         
\centerline{$^{18}$Institut de Physique Nucl\'eaire de Lyon, IN2P3-CNRS,      
                   Universit\'e Claude Bernard, Villeurbanne, France}         
\centerline{$^{19}$RWTH Aachen, III. Physikalisches Institut A,               
                   Aachen, Germany}                                           
\centerline{$^{20}$Universit{\"a}t Bonn, Physikalisches Institut,             
                  Bonn, Germany}                                              
\centerline{$^{21}$Universit{\"a}t Freiburg, Physikalisches Institut,         
                  Freiburg, Germany}                                          
\centerline{$^{22}$Universit{\"a}t Mainz, Institut f{\"u}r Physik,            
                  Mainz, Germany}                                             
\centerline{$^{23}$Ludwig-Maximilians-Universit{\"a}t M{\"u}nchen,            
                   M{\"u}nchen, Germany}                                      
\centerline{$^{24}$Fachbereich Physik, University of Wuppertal,               
                   Wuppertal, Germany}                                        
\centerline{$^{25}$Panjab University, Chandigarh, India}                      
\centerline{$^{26}$Tata Institute of Fundamental Research, Mumbai, India}     
\centerline{$^{27}$University College Dublin, Dublin, Ireland}                
\centerline{$^{28}$Korea Detector Laboratory, Korea University,               
                   Seoul, Korea}                                              
\centerline{$^{29}$CINVESTAV, Mexico City, Mexico}                            
\centerline{$^{30}$FOM-Institute NIKHEF and University of                     
                  Amsterdam/NIKHEF, Amsterdam, The Netherlands}               
\centerline{$^{31}$University of Nijmegen/NIKHEF, Nijmegen, The               
                  Netherlands}                                                
\centerline{$^{32}$Joint Institute for Nuclear Research, Dubna, Russia}       
\centerline{$^{33}$Institute for Theoretical and Experimental Physics,        
                  Moscow, Russia}                                             
\centerline{$^{34}$Moscow State University, Moscow, Russia}                   
\centerline{$^{35}$Institute for High Energy Physics, Protvino, Russia}       
\centerline{$^{36}$Petersburg Nuclear Physics Institute,                      
                   St. Petersburg, Russia}                                    
\centerline{$^{37}$Lund University, Royal Institute of Technology,            
                   Stockholm University, and Uppsala University, Sweden}      
\centerline{$^{38}$Lancaster University, Lancaster, United Kingdom}           
\centerline{$^{39}$Imperial College, London, United Kingdom}                  
\centerline{$^{40}$University of Manchester, Manchester, United Kingdom}      
\centerline{$^{41}$University of Arizona, Tucson, Arizona 85721}              
\centerline{$^{42}$Lawrence Berkeley National Laboratory and University of    
                  California, Berkeley, California 94720}                     
\centerline{$^{43}$California State University, Fresno, California 93740}     
\centerline{$^{44}$University of California, Riverside, California 92521}     
\centerline{$^{45}$Florida State University, Tallahassee, Florida 32306}      
\centerline{$^{46}$Fermi National Accelerator Laboratory, Batavia,            
                   Illinois 60510}                                            
\centerline{$^{47}$University of Illinois at Chicago, Chicago,                
                   Illinois 60607}                                            
\centerline{$^{48}$Northern Illinois University, DeKalb, Illinois 60115}      
\centerline{$^{49}$Northwestern University, Evanston, Illinois 60208}         
\centerline{$^{50}$Indiana University, Bloomington, Indiana 47405}            
\centerline{$^{51}$University of Notre Dame, Notre Dame, Indiana 46556}       
\centerline{$^{52}$Iowa State University, Ames, Iowa 50011}                   
\centerline{$^{53}$University of Kansas, Lawrence, Kansas 66045}              
\centerline{$^{54}$Kansas State University, Manhattan, Kansas 66506}          
\centerline{$^{55}$Louisiana Tech University, Ruston, Louisiana 71272}        
\centerline{$^{56}$University of Maryland, College Park, Maryland 20742}      
\centerline{$^{57}$Boston University, Boston, Massachusetts 02215}            
\centerline{$^{58}$Northeastern University, Boston, Massachusetts 02115}      
\centerline{$^{59}$University of Michigan, Ann Arbor, Michigan 48109}         
\centerline{$^{60}$Michigan State University, East Lansing, Michigan 48824}   
\centerline{$^{61}$University of Mississippi, University, Mississippi 38677}  
\centerline{$^{62}$University of Nebraska, Lincoln, Nebraska 68588}           
\centerline{$^{63}$Princeton University, Princeton, New Jersey 08544}         
\centerline{$^{64}$Columbia University, New York, New York 10027}             
\centerline{$^{65}$University of Rochester, Rochester, New York 14627}        
\centerline{$^{66}$State University of New York, Stony Brook,                 
                   New York 11794}                                            
\centerline{$^{67}$Brookhaven National Laboratory, Upton, New York 11973}     
\centerline{$^{68}$Langston University, Langston, Oklahoma 73050}             
\centerline{$^{69}$University of Oklahoma, Norman, Oklahoma 73019}            
\centerline{$^{70}$Brown University, Providence, Rhode Island 02912}          
\centerline{$^{71}$University of Texas, Arlington, Texas 76019}               
\centerline{$^{72}$Rice University, Houston, Texas 77005}                     
\centerline{$^{73}$University of Virginia, Charlottesville, Virginia 22901}   
\centerline{$^{74}$University of Washington, Seattle, Washington 98195}       
}                                                                             

%% file: acknowledgement_paragraph_r2.tex
%
We thank the staffs at Fermilab and collaborating institutions, 
and acknowledge support from the 
Department of Energy and National Science Foundation (USA),  
Commissariat  \` a l'Energie Atomique and 
CNRS/Institut National de Physique Nucl\'eaire et 
de Physique des Particules (France), 
Ministry of Education and Science, Agency for Atomic 
   Energy and RF President Grants Program (Russia),
CAPES, CNPq, FAPERJ, FAPESP and FUNDUNESP (Brazil),
Departments of Atomic Energy and Science and Technology (India),
Colciencias (Colombia),
CONACyT (Mexico),
KRF (Korea),
CONICET and UBACyT (Argentina),
The Foundation for Fundamental Research on Matter (The Netherlands),
PPARC (United Kingdom),
Ministry of Education (Czech Republic),
Natural Sciences and Engineering Research Council and 
WestGrid Project (Canada),
BMBF (Germany),
A.P.~Sloan Foundation,
Civilian Research and Development Foundation,
Research Corporation,
Texas Advanced Research Program,
and the Alexander von Humboldt Foundation.
%